\documentclass[aps,prl,twocolumn,footinbib]{revtex4}
\usepackage{graphicx, amssymb}

\begin{document}

\title{The shape and motion of a ruck in a rug}

\author{John M. Kolinski, Pascale Aussillous and L. Mahadevan}\email[]{lm@seas.harvard.edu}
\affiliation{School of Engineering and Applied Sciences, Harvard University, Cambridge, MA 02138, USA}
\affiliation{IUSTI CNRS UMR 6595 - Polytech'Marseille - Aix-Marseille Universite,  Marseille cedex 13, France}

\date{\today}

\begin{abstract}
The motion of a ruck in a rug is used as an analogy to explain the role of dislocations in the deformation of crystalline solids. We take the analogy literally and study the shape and motion of a bump, wrinkle or ruck in a thin sheet in partial contact with a rough substrate in a gravitational field. Using a combination of experiments, scaling analysis and numerical solutions of the governing equations, we first quantify the static shape of a ruck on a horizontal plane. When the plane is inclined, the ruck becomes asymmetric and moves by rolling only when the the inclination of the plane reaches a critical angle. We find that the angle at which this first occurs is larger than the angle at which the ruck stops, i.e. static rolling friction is larger than dynamic rolling friction. Once the ruck is in motion, it travels at a constant speed proportional to the sine of the angle of inclination, a result that we rationalize in terms of a simple power balance. We conclude with a simple implication of our study for the onset of rolling motion at soft interfaces. 
\end{abstract}

\maketitle

Since Volterra's first mathematical description of dislocations more than a century ago, the role of these localized defects in the deformation of  crystalline materials is well known \cite{nabarro}. To understand how these dislocations allow for slip along lattice planes, an oft-used analogy, attributed to Orowan \cite{nabarro}, connects the motion of an edge  dislocation and  that of a ruck in a rug. Just as it  is easier to move a rug by having a wrinkle or ruck roll along it rather than by dragging the entire rug, it is easier for a crystal to deform by having a localized defect or dislocation glide or climb rather than sliding globally  along an entire plane. When taken literally, the analog of a ruck in a rug appears in a variety of guises in scientific problems. For  example, at  interfaces between soft and stiff materials, e.g. rubber on glass, motion occurs by the generation and propagation of small wrinkles known as Schallamach waves \cite{schallamach}, not by sliding. In thin films and filaments that interact with a substrate frictionally or adhesively, motion may again occur via the rolling of inchworm-like wrinkles \cite{kendall, gittus}. In flagellar axonemes, individual microtubules often slide relative to each other while propagating microscopic rucks \cite{kamiya}, while in cells, blebs, which are blisters where the cell membrane is detached from the underlying cortex can also propagate in a manner similar to that of rucks \cite{charras}. Finally, earthworms, leeches, and certain worms  move using propagating wrinkles that have been studied and mimicked \cite{trueman,mahadevan}, and surely played a role in creating the analogy in the first place.

\begin{figure}
\includegraphics[scale=0.35]{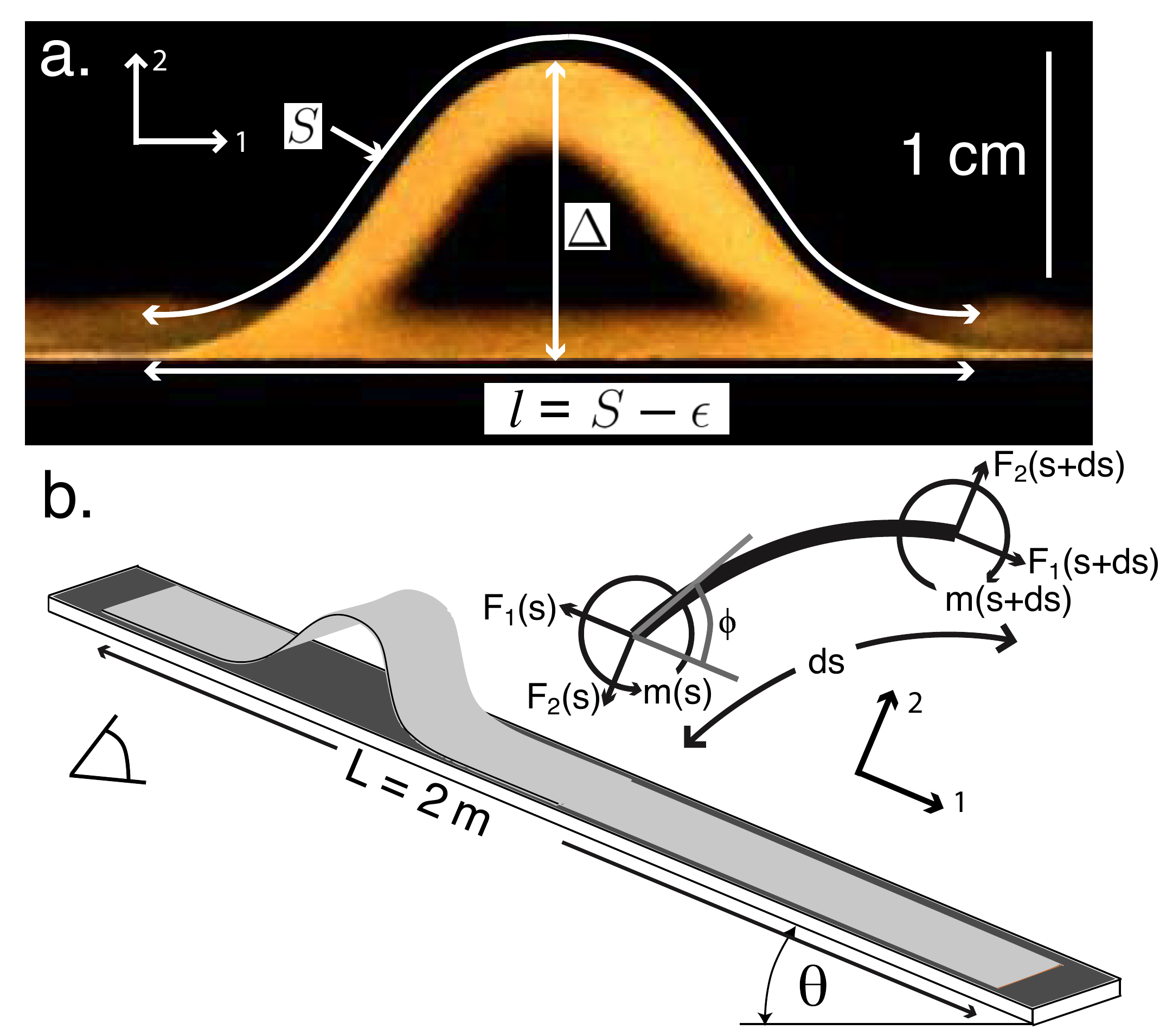}
\caption{(a) A ruck in a rug on a flat substrate. The ruck height is $\Delta$, while its length measured along the arc is $S$ and its horizontal extent $S-\epsilon $, so that $\epsilon$ is analogous to the Burgers vector. The latex rug is $0.25$ mm thick, and the perspective view is a result of the tunnel view through the ruck. (b) The plane is at an angle of inclination, $\theta$. The free-body-diagram in the upper right hand corner shows the local stress and moment resultants for a segment of length $ds$ of the ruck, where $\phi$ is the angle between the local ruck tangent and the plane.   \label{fig.1}}
\end{figure}

Motivated by these examples, here we  study the statics and dynamics of a thin film or filament interacting with a substrate. In Figure~\ref{fig.1} (a) we show the result  of pushing a thin latex sheet (thickness $h=0.25$ mm) lying on a rough plane till it  buckles into a ruck that is stabilized by frictional forces. This ``defect" may be characterized by the difference in the length between the sheet and the projection onto the substrate $\epsilon$, a quantity analogous to the Burgers vector for a dislocation (although it is not a discrete object here); indeed after the ruck has moved through the rug, the rug has been translated by a distance $\epsilon$. When the plane on which the ruck sits is tilted, the ruck remains stationary until the inclination of  the substrate reaches a critical value, and only then does the ruck move, and then does so at a constant  velocity, shown schematically in Figure~\ref{fig.1} (b). These simple observations raise a number of questions associated with the shape and dynamics of a thin sheet in partial contact with a rough substrate, which we now address.

\begin{figure}
\includegraphics[scale=0.45]{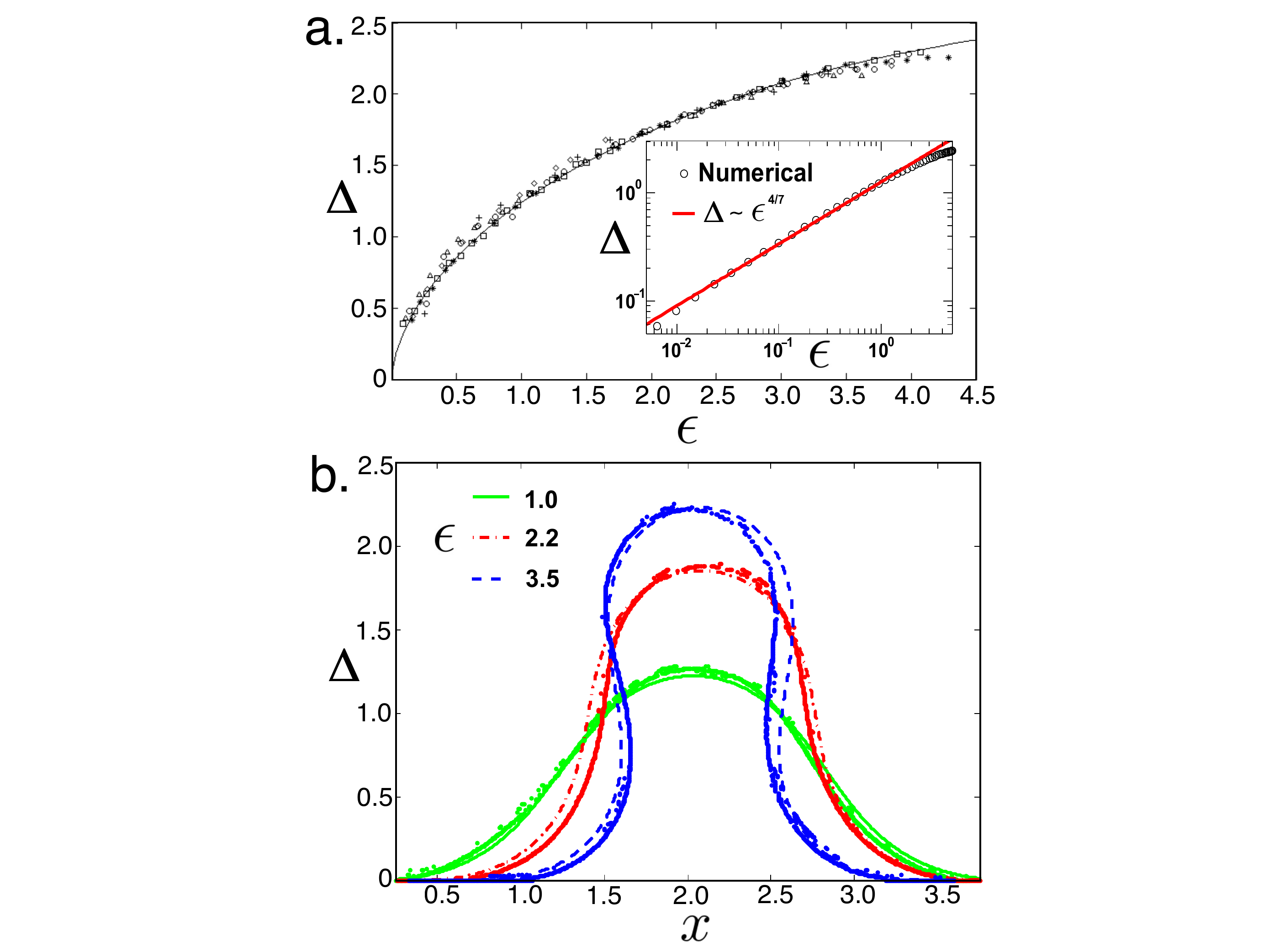}
\caption{(a) The scaled ruck height, $\Delta$ as a function of the excess length $\epsilon$, for different materials and thicknesses, h, (0.25 mm (+), 0.5 mm ($\diamond$), 0.75 mm ($\bigtriangleup$), 1.0 mm (*), 1.5 mm (o)) and for the numerically calculated solutions (solid line) of ruck shapes collapses on to a single curve. All lengths are normalized by the elastic gravity length, $\ell_g = ({\frac{B}{\rho g h }})^{\frac{1}{3}}$. The inset shows the numerically calculated values (o) of $\Delta$ as a function of $\epsilon$  on a log-log plot and compares well with the scaling law $\Delta \sim \epsilon^{\frac{4}{7}}$ given by (4). (b) Ruck shape as a function of the excess length, $\epsilon$; solid lines correspond to experimental measurements, while the dashed lines correspond to numerical solutions of (1)-(3) \label{fig.2}}
\end{figure}

Assuming the ruck to be inextensible, a good approximation when the latex sheet is compressively loaded, implies the kinematic relations between the instantaneous position of material points in the ruck  $x(s,t), y(s,t)$ and the orientation of the local tangent vector $\phi(s,t)$ relative to the plane on which the ruck sits, parametrized in terms of  the arc length coordinate $s$ 
\begin{eqnarray} 
\label{kin} x_{,s} = \cos \phi,  ~~ y_{,s} &=& \sin \phi. \end{eqnarray} 
Here $\partial A /\partial a \equiv A_{,a}$. Force balance parallel 
and perpendicular to the plane yields the equations of motion (see Figure~\ref{fig.1} (b) for a pictorial derivation):
\begin{eqnarray}
\label{balance}   F_{1,s} + \rho gh \sin \theta  + f_1 &=&  0, \nonumber \\ 
 F_{2,s} - \rho gh \cos \theta +f_2 &=&  0 \nonumber \\ 
  M_{,s} - F_{1}\sin \phi  + F_{2} \cos \phi  + m &=&  0.
 \end{eqnarray}  
Here  $F_{1}$ and $F_{2}$ are the integrated stress resultants in the film of density $\rho$,  $M$ is the torque resultant and $g$ is the gravitational  acceleration, and $f_1,f_2, m$ are the volumetric forces and torque at a cross-section (and include any potential contributions from inertia or other body forces).  Closure of the system of equations requires a specification of the torque $M$. Latex rubber is a viscoelastic material which we model as a simple Voigt solid, so that we can write  $M = E I_a \phi_{,s}+\mu I_a \phi_{,st}$, where $I_a$ is the area moment of inertia of the cross-section, $E$ is the elastic modulus and $\mu$ is the viscosity of the material. Finding the shape of either the static or dynamic ruck requires the specification of boundary conditions at the two contact lines that demarcate the locations where the ruck leaves and regains contact with the substrate, and read
\begin{eqnarray} 
\label{bcs} x(0) = y(0) = \phi (0) = \phi_{,s}(0) =  y(S) =  \phi (S) =  0. 
\end{eqnarray} 
Here $S$ is the length of the ruck that is free, and we have chosen the origin of the ruck to be the left contact line. The condition of the curvature vanishing at either end follows from the fact that there are no localized torques at the contact lines \footnote{ The condition $\phi_{,s}(S) = 0$ is automatically satisfied as can be verified by multiplying the last equation in (\ref{balance}) by $\phi_s$, and integrating the result in light of the other two conditions. }.

For small amplitude rucks, the excess length  $\epsilon \ll S$, and furthermore, the projected length of the ruck $l \sim S$. Then, geometry implies that the ruck height $\Delta = \sqrt{\epsilon l}$, and its curvature $\kappa \sim \Delta/l^2$. Comparing the elastic bending energy, $U_e \sim EI \kappa^2 l$, with the gravitational potential energy, $U_g \sim \rho gh l \Delta$, together with the geometric  relations $\kappa \sim \Delta/l^2, I \sim h^3$ yields the dimensional scaling laws $l \sim \epsilon^{1/7} (Eh^2/\rho g)^{2/7}, \Delta \sim \epsilon^{4/7} (Eh^2/\rho g)^{1/7} $, and the  dimensional energy per unit length required to form a ruck scales as $\rho g h \ell_g^{9/7} \epsilon^{5/7}$. Using the elastic gravity length, $\ell_g = (\frac{Eh^2}{\rho  g})^{\frac{1}{3}}$ as a natural length scale in the problem then allows us to write the dimensionless length and height of the ruck as
\begin{eqnarray} 
l/\ell_g \sim \epsilon^{\frac{1}{7}}, \Delta/\ell_g  \sim \epsilon^{\frac{4}{7}}. \label{scale}
\end{eqnarray} 
For rucks of large amplitude, we must solve the free boundary problem (\ref{kin}-\ref{bcs}) in the static limit (with $ f_1=f_2=m=0$), after normalizing all lengths by the elastic gravity length, $\ell_g = (\frac{EI_a}{\rho h g})^{\frac{1}{3}}$, stresses by $N = \rho g h \ell_g $, with the dimensionless variables $\hat x =x /\ell_g, \hat{F_1} = F_1/N, \hat{\epsilon} =\epsilon/\ell_g $, etc., and then drop the hats unless otherwise noted.  In Figure~\ref{fig.2} (a), we show the dimensionless height $\Delta$  as a function of the dimensionless excess length $\epsilon$, obtained by solving  (\ref{kin}-\ref{bcs}) using a shooting method implemented in MATLAB.  We also plot the experimentally determined height of the ruck for various values of $\epsilon$ and see that the experimental and simulation data collapse onto a single curve. In the inset,  we plot the same data on a log-log plot and see that our scaling law (\ref{scale}) works for small to moderate values of $\epsilon$. In  Figure~ \ref{fig.2} (b), we show the shape of the ruck for various values of $\epsilon$, determined experimentally and numerically, and see that they compare well \footnote{There is a critical value of the excess length $\epsilon_c$ above  which the ruck loses 
symmetry and tilts to one side, a question that we will not address further here.}.
 
\begin{figure}
\includegraphics[scale=0.65]{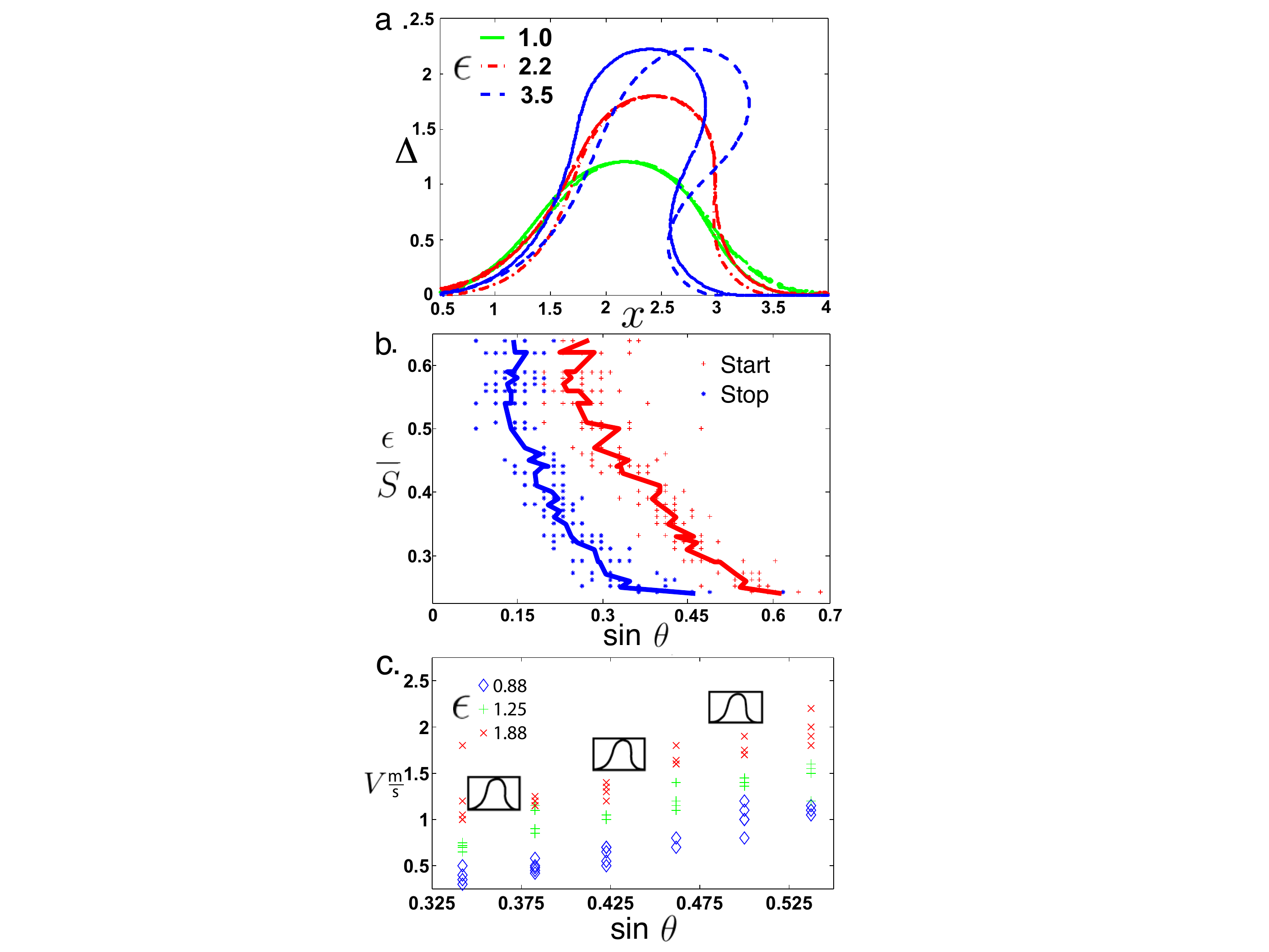}
\caption{(a) The ruck shape  on an inclined plane with $\sin \theta =0.3$ shows the increasing asymmetry as  $\epsilon$ increases. Solid lines correspond to numerical solutions of (1)-(3), while the dashed lines correspond to experiments. (b) The ratio of the excess length to the contour length $\epsilon/S$ vs. $\sin \theta$ at the onset of motion  with $\theta=\theta_g$ and when motion stops with $\theta=\theta_s <\theta_g$, i.e. dynamic rolling friction is less than static rolling friction (c) Ruck speed $V$ vs. $\sin\theta$  when $\theta >\theta_s$ for three values of $\epsilon$, 0.88,1.25 and 1.88,  is consistent with the scaling law $V \sim \sin \theta$ determined by the balance between visco-elastic power dissipation in the latex and gravitational power input (see text). All experiments were done with a latex sheet of thickness $h=0.75$ mm, and the shapes of the ruck  (inset) correspond to $\theta= 20^0, 25^0, 30^0$, $\epsilon = 1.88$. \label{fig.3}}
\end{figure}

When the plane on which the ruck is formed is tilted, the ruck becomes asymmetric, but does not move. Since only the gravitational energy scales differently, following our earlier argument yields the scaling laws for the dimensionless length of the ruck $l/\ell_g \sim \epsilon^{1/7} (\cos  \theta)^{-2/7}$ and its dimensionless height $\Delta/\ell_g \sim \epsilon^{4/7} (\cos \theta)^{-1/7}$, while the dimensional energy required to create a ruck on a tilt scales as $\rho g h \ell_g^{9/7} \epsilon^{5/7} ( \cos \theta)^{4/7}$. To go beyond scaling arguments, we solve the equations (1)-(3) for various values of the tilt $\theta$ and excess length $\epsilon$ leading to the results shown in Figure~ \ref{fig.3} (a) along with the experimentally observed shapes, which compare well for moderate values of $\epsilon$ but show substantial deviations for large values of $\epsilon$.
 
\begin{figure}
\includegraphics[scale=0.65]{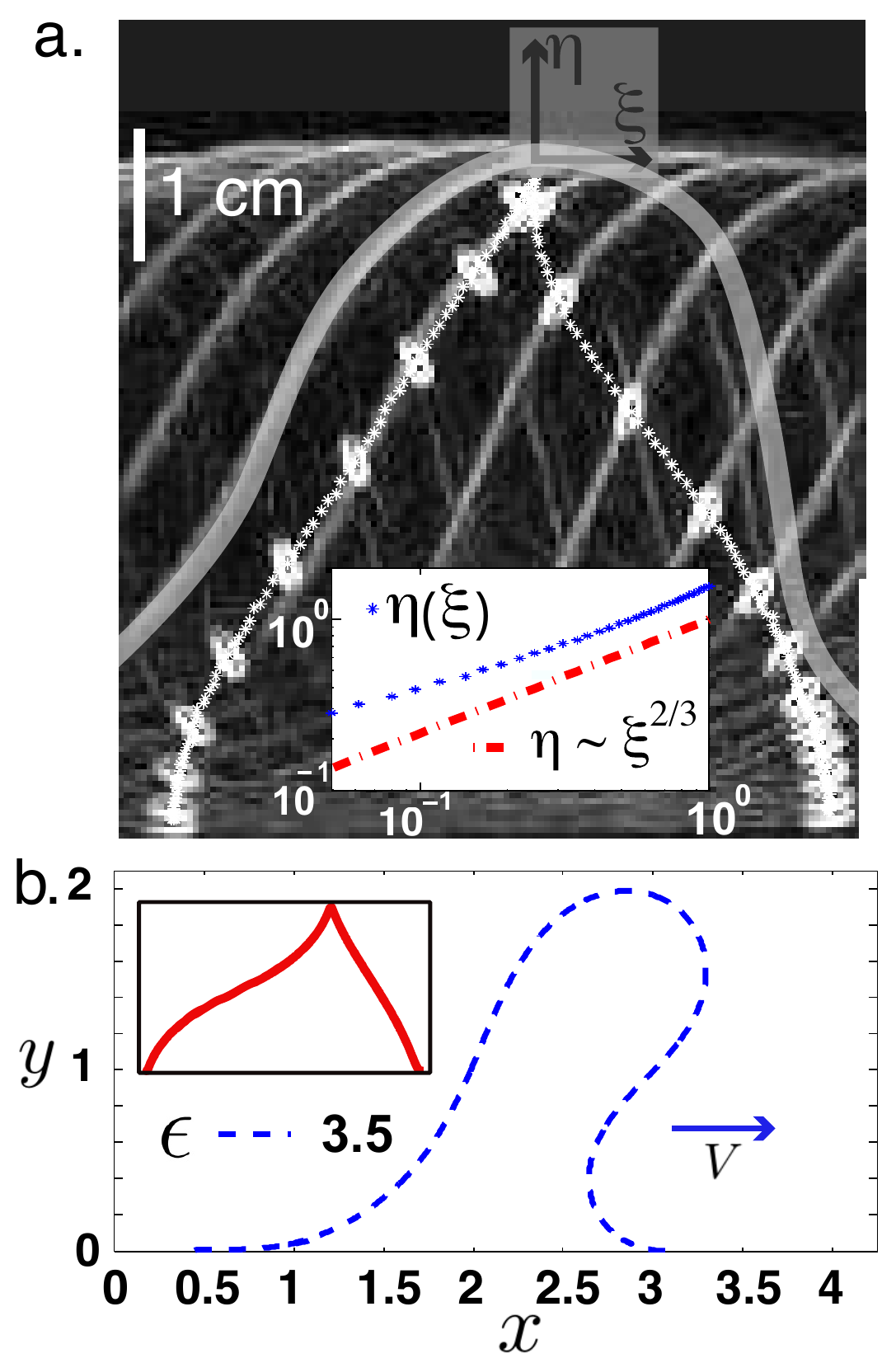}
\caption{(a) The path of a point on the rug as a ruck moves it is plotted in white (*)  on top of the time-lapse images of the ruck.  In the inset is a plot of a point on the ruck as it approaches its zenith (*), compared  the cuspidal curve $y \sim x^{2/3}$ that characterizes the kinematics of a point on a rigid wheel rolling on a rigid substrate. (b) For a slowly moving ruck, the dynamic shape of a ruck is close to its  static shape (dashed line) along with the particle path of a point on it (inset, solid line). The experimental curves correspond to $\epsilon=1.25$, $\theta=20^o$, $V=0.7$ m/s while the numerical curves correspond to $\epsilon = 3.5$, $\theta = 17^o$. \label{fig.4}} \end{figure}

When the angle of inclination is larger than a critical threshold, the ruck begins to move. It does so by rolling rather than sliding. Measuring $\epsilon$ at the beginning of the  run to that at its end,  we find that it is conserved to within $2\%$. Starting with a stationary ruck in a rug on an incline, and increasing  the angle of inclination slowly until it starts to move allows us to determine the critical angle $\theta_g$ at which rolling starts,. Once the ruck is in steady motion at a speed $V$, the incline is untilted slowly (relative to the movement of the ruck, i.e. $\theta_{,t} \ll V/S$) to determine the angle $\theta_s$ at which the ruck stops. In Figure~\ref{fig.3} (b), we see the marked difference between static and dynamic rolling  friction characterized in terms of $\theta_g$ and $\theta_s$; $\theta_g$  is an indicator of a critical torque  which a ruck (of  given shape and size) must overcome in order to begin rolling (a measure of static rolling friction), and  $\theta_s <\theta_g$ measures the dynamic rolling friction resistance, analogous to well known, but still incompletely understood difference between static and dynamic sliding friction. The existence of a critical inclination for the onset of rolling is consistent with the existence of a critical torque due to microscopic interactions at the contact line.  Interestingly, there is a similar threshold for the motion of  dislocations \cite{nabarro}, but we will not pursue the possible mechanisms

When $\theta > \theta_g$, the ruck rolls down the inclined plane, quickly reaching a steady speed $V$ and a steady shape that is similar to its static shape, at least for small velocities. In Figure~\ref{fig.3} (c), we show the variation of $V$ with the inclination $\theta$ for different values of $\epsilon$ and observe that $V \sim \sin \theta$. Since the ruck moves by rolling rather than sliding, the gravitational power must be balanced by either air drag or dissipation within the ruck. For small amplitude rucks, balancing the viscoelastic power dissipation per unit length $\mu h^3 ( V \Delta /l^3)^2 l$ \footnote{Here we have used the relation that the power dissipated is $\mu h (h \kappa_t)^2 l$, with $\kappa_t \sim V \kappa_s \sim V \Delta /l^3$.} with gravitational power input per unit length $\rho h l V g \sin \theta $, we find that the dimensional speed $V \sim \rho g \epsilon^{-2/7} l_g^{30/7} \sin \theta/\mu h^2 $, i.e. smaller rucks travel faster than  bigger rucks, with a speed proportional to the  sine of the angle of the incline \footnote{For small amplitude fast moving rucks, where the resistance arises from air drag, balancing the gravitational power $\rho h l V g \sin \theta$ with the power to overcome air drag $\rho_f V^3 \Delta^2$ yields $ V \sim (\rho gh \sin \theta/\rho_f)^{1/2} (\ell_g/\epsilon)^{3/14}$.}. While the latter observation is consistent with our observations as shown in Fig. 3 (b), we do not see evidence that smaller rucks are faster than larger ones. We attribute this to the fact that experimentally, small rucks do not move until the angle of inclination is so large that a large part of the sheet slips before the ruck moves. Thus, we are experimentally limited to studying a range of relatively large values of $\epsilon$ when the shape of the ruck does not show the simple scaling behavior that led to this conclusion.
 
To complete our investigation of the rolling  ruck, we compare it to the rolling of a rigid wheel on a rigid plane, where the particle paths are simple cycloidal arcs separated by cusps. In Figure~\ref{fig.4} (a), we show the path of a particle in the rug that is transported by the ruck.  In the inset of Figure~\ref{fig.4} (a) we plot the local variation of the trajectory of a material point in the ruck in the neighborhood of the cusp,  consistent with the curve $ \eta \sim \xi^{2/3}$ that characterizes the cuspidal profile for a particle on a cycloidal trajectory. In Figure~\ref{fig.4} (b), for comparison, we show the particle path using a numerically calculated shape of the stationary ruck. We see that while particle paths in a rolling ruck are similar to those in a rolling wheel in the vicinity of three points (when it starts, stops and near its zenith), in general the rolling wheel and rolling ruck are  quite different, given the deformable nature of the latter. 

Our study has answered some basic questions about the shape and motion of a soft ruck in a rug, but leaves us with  the question of understanding the transition from statics to dynamics in the ruck. Here, we wish to simply point out that this requires us to account forthe inclusion of the effect of frictional interactions at the elastic contact lines. Then the contact line torque  $EI\phi_{,s}(0) \ne EI \phi_{,s}(S) \ne 0$ in general, so that global torque balance on the ruck then implies $M_r= EI\phi_{,s}(0) - EI \phi_{,s} (S) + \rho g Sl p(\epsilon/S,\theta) \ne 0$, where $p(.,.)$ is a dimensionless function that characterizes the shape of the ruck, and the inclination of the plane. At the onset of rolling, $M_r/\rho ghSl=M^*(\theta, \epsilon/S,...) =M^*$, a scaled threshold torque. Determining $M^*$ in terms of the various interactions at the contact line constitutes a basic question in the dynamics of soft rolling and indeed in the context of all the applications mentioned in the introduction. However, just as phenomenologically derived  velocity weakening laws  allow us to make progress on certain sliding friction problems by circumventing  the question of the onset and stoppage of sliding at interfaces, for rolling at interfaces one might be also able to do the same, given that we now have experimental evidence for the existence of a similar scenario.
 
{\bf Acknowledgment} As this work, begun six years ago in Cambridge, was being completed, we became aware of very recent work addressing the inertial dynamics of the motion of a ruck by  D. Vella, A. Boudaoud and M. Adda Bedia, who we thank for their preprint.

\end{document}